\documentclass[reprint,prl,floatfix]{revtex4-2}

\usepackage{amsmath}
\usepackage[T1]{fontenc}
\usepackage[latin1]{inputenc}
\usepackage{graphicx}
\usepackage{amssymb}
\usepackage{dcolumn}
\usepackage{color}
\usepackage{bm}
\usepackage{multirow}
\usepackage[normalem]{ulem}
\usepackage{url}
\usepackage[ colorlinks = true,
            linkcolor = red,
            urlcolor = red,
            citecolor = blue,
            anchorcolor = red]{hyperref}
\usepackage[colorinlistoftodos]{todonotes}
\usepackage{bibentry}

\usepackage{lineno}
\usepackage{pdfpages}
\makeatletter
\AtBeginDocument{\let\LS@rot\@undefined}
\makeatother

\begin{document}
\title{Topological Thermal Hall Conductance of Even Denominator Fractional States}

\author{Arup Kumar Paul$^{1, \dagger}$, Priya Tiwari$^{1, \dagger}$, Ron Melcer$^{2}$, Vladimir Umansky$^{1}$, and Moty Heiblum$^{1,*}$}

\affiliation{$^{1}$ Braun Center for Submicron Research, Department of Condensed Matter Physics, Weizmann Institute of Science, Rehovot, Israel.\\
 $^{2}$ Qedma Quantum Computing, Tel Aviv, Israel.}

\begin{abstract}
The even denominator fractional quantum Hall (FQH) states $\nu=5/2$ and $\nu=7/2$ have been long predicted to host non-abelian quasiparticles (QPs). Their present energy-carrying neutral modes are hidden from customary conductance measurements and thus motivate thermal transport measurements, which are sensitive to all energy-carrying modes. While past `two-terminal' thermal conductance ($k_{2t}T$) measurements already proved the non-Abelian nature of the $\nu=5/2$ FQH state, they might have been prone to a lack of thermal equilibration among the counter-propagating edge modes. Here, we report a novel thermal Hall conductance measurement of the $\nu=5/2$ and $\nu=7/2$ states, being insensitive to equilibration among edge modes. We verify the state's non-Abelian nature, with both states supporting a single upstream Majorana edge mode (hence, a particle-hole Pfaffian order). While current numerical works predict a different topological order, this contribution should motivate further theoretical work.
\end{abstract}

\maketitle
\section{Introduction}

The search for quantum states that host non-abelian quasiparticles (QPs), localized or propagating, intensified in the past few years, stemming from their unique characteristics and potential to serve as robust qubits in a noisy environment. Among several proposed implementations in the quantum Hall (QH) effect regime, even-denominator fractional states, such as the $\nu=5/2$ and $\nu=7/2$, are the leading candidates~\cite{willett1987observation,moore1991nonabelions,morf1998transition,levin2007particle,lee2007particle,nayak2008non,stern2010non,storni2010fractional,wojs2010landau,zucker2016stabilization,rezayi2017landau,antonic2018paired} with numerical calculations predicting an anti-Pfaffian (A-Pf) topological order~\cite{morf1998transition,levin2007particle}. This order supports a fractional downstream `charged mode' (in addition to downstream integer modes) and three upstream Majorana modes~\cite{moore1991nonabelions,stern2010non,storni2010fractional,wojs2010landau,zucker2016stabilization,rezayi2017landau,antonic2018paired,pan1999strongly,willett2009measurement,liu2011evolution,feldman2021fractional,willett2023interference}.

The $\nu=5/2$ state has been studied extensively in the experimental realm. Earlier tunneling measurements pointed at various possible states, such as the A-Pf order~\cite{zucker2016stabilization,radu2008quasi,lin2012measurements} or different abelian and non-abelian orders~\cite{zucker2016stabilization,baer2014experimental}. However, more recent heat transport and shot noise measurements observed a `particle-hole Pfaffian' (PH-Pf) order~\cite{banerjee2018observation,dutta2022isolated,dutta2022distinguishing}. The published works include measurements of (i) the two-terminal thermal conductance coefficient $k_{2t}$ for all participating edge modes (two integers and fractions)~\cite{banerjee2018observation}; (ii) $k_{2t}$ for the isolated fractional channel (fractional charged and Majorana)~\cite{dutta2022isolated}; and (iii) shot noise, testing the chirality of the isolated fractional channel~\cite{dutta2022distinguishing}. The observed PH-Pf order differs from the A-Pf by supporting a single upstream Majorana mode (central charge, $c=-1/2$)~\cite{banerjee2018observation,dutta2022isolated}. The less explored $\nu=7/2$ state is also expected to be non-abelian~\cite{pan1999strongly,willett2009measurement,liu2011evolution,feldman2021fractional,willett2023interference}; however, its topological order was not established.

The `two-terminal' thermal conductance measurement~\cite{jezouin2013quantum,srivastav2019universal} relies on a small floating ohmic contact (`source'), 
with a known power being dissipated in it. The equilibrium temperature ($T_H$) of the source and $k_{2t}$ are related to the net power ($J_{2t}$) that leaves the contact by,\begin{equation} 
J_{2t}=0.5k_{2t}(T_{H}^2-T_{0}^2) 
\end{equation} 
where $T_0$ is the ground temperature. For a single ballistic chiral abelian mode, the thermal conductance ($G_{th}=dJ_{2t}/dT_H$) is quantized to $k_{2t}T=k_0T$, with $k_0=\pi^2k_B^2/3h$, $T$ temperature, and $k_B$ and $h$ are Boltzmann and Planck constants, respectively~\cite{kane1997quantized,cappelli2002thermal}. In quantum Hall states featuring both downstream and upstream edge modes, each with respective thermal conductances $k_dT$ and $k_uT$, the net thermally equilibrated thermal conductance is $k_{2t}T=(k_d-k_u)T$. However, if the edge modes do not interact, then $k_{2t}T=(k_d+k_u)T$. Consequently, partial thermal equilibration can lead to any value between the two extremes of $k_{2t}$~\cite{srivastav2021vanishing,srivastav2022determination}. For example, in the $\nu=5/2$ state, if an existing Majorana mode (with thermal conductance $k_0T/2$~\cite{lee2007particle,nomura2012cross,sumiyoshi2013quantum,do2017majorana,kasahara2018majorana,yokoi2021half}) will not interact, the two-terminal thermal conductance would increase by $k_0T$. Such a lack of equilibration might create the impression of the fully equilibrated PH-Pf order ($k_{2t}=2.5k_0$)~\cite{read2000paired,simon2020energetics,simon2018interpretation,feldman2018comment,simon2020partial,hein2023thermal,manna2022full}, while the true order is A-Pf (with equilibrated $k_{2t}=1.5k_0$).

To mitigate the ambiguity arising from partial or no equilibration, we utilize a new measurement configuration that directly determines the topological thermal Hall conductance coefficient $k_{xy}$, of the $\nu=5/2$ and $\nu=7/2$ states~\cite{melcer2023direct,melcer2024heat}. This method involves separately measuring downstream and upstream heat flows ($J_{down}$ and $J_{up}$) to determine the coefficients $k_d$ and $k_u$, ultimately yielding $k_{xy}=k_d-k_u$, which represents the topological order of the states~\cite{melcer2023direct,melcer2024heat}.


\section{Device and setup} 
\begin{figure}[ht]
\includegraphics[width=0.48\textwidth]{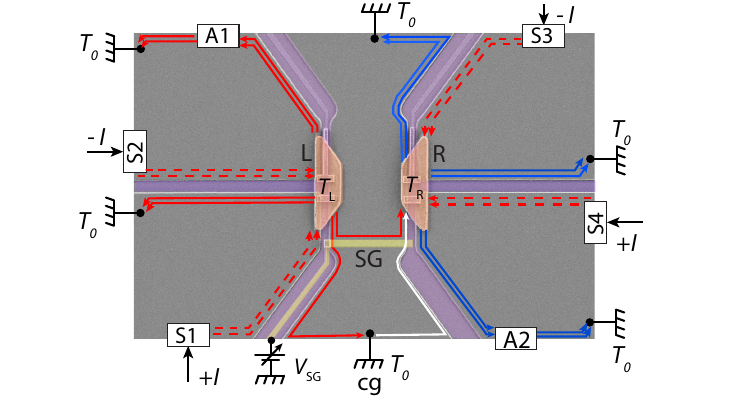}
\caption{\label{fig1}\textbf{The device and measurement setup :} An SEM image of the device showing the floating contacts (orange), \textbf{L} (with temperature $T_L$), and \textbf{R} (with temperature $T_R$), the continuous side gate, \textbf{SG} (yellow, charged by $V_{SG}$), and the mesa arms (grey) defined by etching grooves (purple) in the GaAs. The arrowed lines represent the QH edge modes. Red dashed arrows show current-carrying edge modes (injected from contacts \textbf{S1} and \textbf{S2} or \textbf{S3} and \textbf{S4}) that heat the floating contacts. Solid red and blue arrows represent the outgoing edge modes from contacts \textbf{L} and \textbf{R}, respectively, and they terminate at cold ground (cg) contacts connected to the dilution mixing chamber plate at base temperature $T_0$. The white arrow represents edge modes originating from cg. \textbf{A1} and \textbf{A2} are amplifier contacts. 
}
\end{figure}

The devices under test (shown in a false color SEM image in Fig.~\ref{fig1}) were fabricated in a two-dimensional electron gas (2DEG) confined in a GaAs-AlGaAs heterostructure, with short-period superlattice (SPSL) type doping~\cite{umansky2009mbe,umansky2013mbe}. The 2DEG mobility $\sim 9.1\times 10^6$~cm$^2$/V.s and density $\sim2.9\times 10^{11}$~cm$^{-2}$, are measured at 4.2~K in the dark.

The QH responses around the two states of interest, measured in a Hall bar, are shown in \textbf{Supplemental Material-1} (\textbf{SM-1}). The tested devices (Fig.~\ref{fig1}) consist of two identical floating ohmic contacts, denoted by L (left) and R (right), each with an area of $\sim49$~$\mu$m$^2$. Both contacts are covered with $\sim25$~nm thick dielectric HfO$_2$ followed by a grounded thin metallic layer, which increases the contact capacitance and thus suppresses the charging energy~\cite{slobodeniuk2013equilibration,sivre2018heat}. The 2DEG (grey) under each floating contact is grooved (purple), thus `forcing' the incident current to enter the contacts~\cite{melcer2023direct,melcer2024heat}. Contacts {L} and {R} are separated by 10 $\mu$m (or 30 $\mu$m) of the intermediate 2DEG bulk. A side gate ({SG}, at the lower part of the intermediate bulk), when `pinched', guides the edge modes from {L} to {R}. When the SG is not biased, the chiral modes flow to the ground. Partial biasing of the SG directs chosen edge modes to the ground. Contacts L and {R} are also attached to two separate long mesa arms ($\sim $ 120 $\mu$m long each), with two current sourcing contacts ({S1} and {S2} or {S3} and {S4}), amplifier contacts ({A1} and {A2}), and `cold grounds' (cg) maintained at base temperature $T_0$ (Fig.~\ref{fig1}). The amplifiers amplify the Johnson-Nyquist (J-N) noise carried by edge modes (solid red and blue arrows) emanating from the floating contacts, facilitating the determination of the contact's temperature (refer to \textbf{SM-2} for details)~\cite{melcer2023direct}.

Each of the floating contacts is heated with a known power supplied by equal and opposite DC currents ($+I$ and $-I$) emanating from contacts {S1} and {S2} (for contact {L}) or {S3} and {S4} (for contact {R}). Each floating contact's potential remains zero, thus eliminating possible emanating shot noise and assuring that the outgoing edge modes carry only the J-N noise. Under these conditions, the dissipating power is $P_d=I^2R$, with $R=h/\nu e^2$ the QH resistance. During measurements, one of the floating contacts is heated and acts as the temperature source (S) (see \textbf{SM~Fig.~2}). The second floating contact is heated by the arrival power (carried by the edge modes and the bulk), thus acting as a power meter (PM). The increased PM temperature is `translated' (via a calibration process) to the arrival power from S (see \textbf{SM~Fig.~3})~\cite{melcer2023direct,melcer2024heat}. In the `downstream configuration', the floating contact L plays the role of the source while the floating contact R plays the PM role. In the `upstream configuration', the functions of L and R are interchanged.

We now briefly review the method of obtaining the topological thermal Hall conductance. The arriving downstream heat flow to the PM, $J_{down}$, and the exiting heat from it, $J_{out}=J_e +J_{\gamma}$ ($J_e$ - via edge modes, and $J_{\gamma}$ via phonons) leads to net dissipated power in the PM with an equilibrium temperature $T_{PM}$. This temperature is `converted back' to the arriving power, $J_{down}$, via a separate heating process by a known power (as described above and also in \textbf{SM-2}). The same procedure is applied when the upstream heat flow, $J_{up}$, is measured by interchanging the heated and the absorbing contacts (L and R).
 The topological thermal conductance coefficient $k_{xy}$ is determined from $J_{down}$ and $J_{up}$, using,\begin{equation}
J_{down}-J_{up}=\frac{(k_{d}-k_{u})}{2} (T^2-T_0^2)=\frac{k_{xy}}{2} (T^2-T_0^2)
\end{equation}
where $T$ is the source (contact L for downstream and contact R for upstream) temperature~\cite{melcer2023direct}. It is important to stress that thermal equilibration between modes, or lack of it, as well as the contributions of the bulk or due to edge-reconstruction, are eliminated by the subtraction process of $J_{down}-J_{up}$~\cite{melcer2023direct,melcer2024heat}. However, the energy loss of the `down' and `up' propagating modes, not necessarily being equal, can not be recovered. Hence, the L-R separation is kept short.

\section{Results} 
\begin{figure}[ht]
\includegraphics[width=0.48\textwidth]{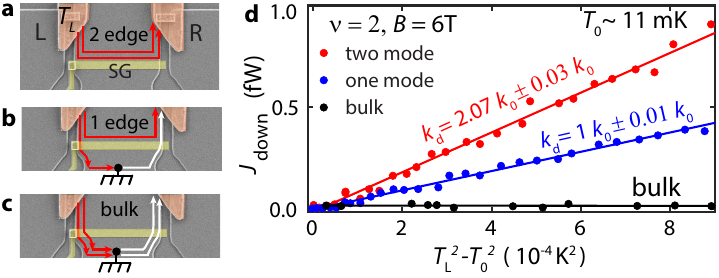}
\caption{\label{fig2}\textbf{Thermal conductance of $\nu=2$:} For downstream heat flow, contact \textbf{L} acts as the source, and contact \textbf{R} acts as the PM. \textbf{(a)} Configuration for downstream heat flow measurement with two edge modes (red arrows) carrying heat from \textbf{L} to \textbf{R} ($V_{SG}=-2$~V). \textbf{(b)} Configuration for the inner edge mode reaching contact \textbf{R}, while the outer edge mode is diverted to ground ($V_{SG}=-0.85$~V). \textbf{(c)} Configuration for bulk heat conduction with all the edge modes diverted to the ground ($V_{SG}=0$~V). The white arrows represent edge modes starting from the ground.
\textbf{(d)} Downstream heat flow $J_{down}$ arrives at contact R and heats the contact. The $J_{down}$ is plotted as a function of $T_L^2-T_0^2$ ($T_0\sim11$~mK, electron temperature). The red, blue, and black colors correspond to heat conduction by two modes, one mode, and only the bulk, respectively. Solid lines are the linear fits to the data. 
} 
\end{figure}
We first test the configuration at the well-understood filling $\nu=2$, performing the measurements at the conductance plateau's center (B=6~T) on a device with an L-R separation of 10 $\mu$m. The Fig.~\ref{fig2}(a) shows the measurement configuration with a fully pinched SG gate ($V_{SG}=-2$~V), with two downstream edge modes leaving contact L (source) and reaching contact R (PM). Increasing the gate voltage to $V_{SG}=-0.85$~V (and to $0$~V) enables the outer (and also the inner) modes to reach the ground as shown in Fig.~\ref{fig2}(b) (and Fig.~\ref{fig2}(c)). \textbf{SM-3} shows gate transmittance as a function of $V_{SG}$ for the studied filling factors. Fig.~\ref{fig2}(d) shows the dependence of the (calibrated) incoming power $J_{down}$ (reaching contact R) as a function of $T_L^2-T_0^2$ ($T_0\sim 11$ mK), where $T_L$ is the temperature of contact L. The red, blue, and black dots correspond to downstream heat flow by two modes plus the bulk, one mode plus the bulk, and only the bulk ($V_{SG}=0$~V), respectively. With \textbf{Eq. (1)}, we find $k_d=2.07k_0\pm0.03k_0$ for two propagating edge modes and $k_d=1k_0\pm0.01k_0$ for a single edge mode. The heat flow through the bulk and the upstream heat flow are both negligibly small; hence, $k_{xy}(\nu=2)\approxeq2k_0$, as expected. 
To assess the impact of bulk's heat conductance, we also measured $k_{xy}$, away from the center of the $\nu=2$ plateau (at 5.6~T) (see \textbf{SM-4}). Here, despite having finite bulk heat conduction, we find similar values for $k_{xy}$, thus reassuring our method's effectiveness in determining the topological order of the states.

\begin{figure}[ht]
\includegraphics[width=0.48\textwidth]{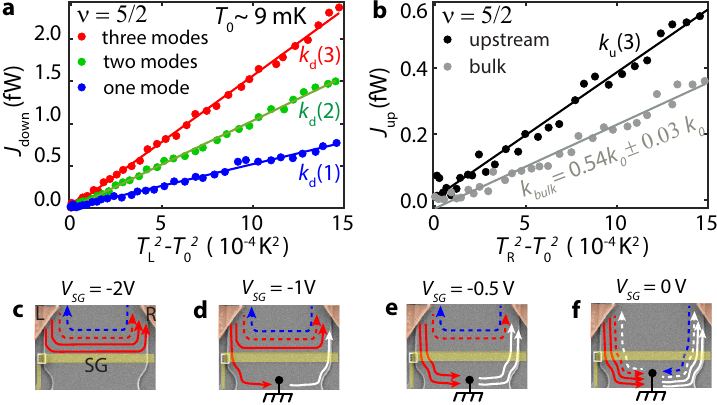}%
\caption{\label{fig3}\textbf{Thermal conductance of $\nu=5/2$:} \textbf{(a)} Downstream heat flow $J_{down}$ that leaves the heated contact \textbf{L} and reaches contact \textbf{R} vs. $T_L^2-T_0^2$ ($T_0\sim9$~mK, electron temperature). The red, green, and blue data points are for all edge modes (two integers and the innermost fractional mode), two inner modes (the inner integer mode and the fractional mode), and only the fractional mode, respectively. The downstream heat conductance coefficients $k_d(N=3,2,1)$ are determined from the linear fittings shown by the solid lines and are given in Table.~\ref{Table1}. \textbf{(b)} Upstream heat flow $J_{up}$ measured by changing the roles of L and R with all modes participating (for fully pinched gate SG, $N=3$) is plotted as a function of $T_R^2-T_0^2$ (in black). The grey dots are for bulk heat transport when SG is fully open. \textbf{(c)-(f)} Different configurations of edge modes propagation between L and R: \textbf{(c)} Fully pinched \textbf{SG} ($V_{SG}=-2$~V), \textbf{(d)} and \textbf{(e)} Partially pinched SG with one ($V_{SG}=-1$~V) and two ($V_{SG}=-0.52$~V) edge modes diverted to ground. \textbf{(f)} Fully open SG ($V_{SG}=0$~V). The arrows indicate edge modes: solid red for downstream integer modes, dashed red for downstream fractional mode, and blue dashed arrow for upstream modes. White arrows show edge modes emanating from the ground.}
\end{figure}

Having established the method's effectiveness, we now focus on the even denominator states. Starting with the $\nu=5/2$ state, Fig.~\ref{fig3}(a) shows the $J_{down}$ vs. $T_L^2-T_0^2$ ($T_0\sim9$~mK) plots measured at the center of $\nu=5/2$ plateau ($\mathrm{B}\sim4.64$~T) for different SG voltages (refer to \textbf{SM-3}). Fig. \ref{fig3}(b) shows the $J_{up}$ vs. $T_R^2-T_0^2$ plots for fully pinched SG (all edge mode participating) and of the bulk conduction when SG is fully transparent (see \textbf{SM-5} for partially closed SG). Here, $T_R$ is the temperature of contact R. Fig.~\ref{fig3}(c) to \ref{fig3}(f) shows the edge mode configurations between the two floating contacts with different SG voltages. The solid and dashed red arrows in these figures represent the downstream integer and the $1/2$ fractional-edge mode, respectively. The blue arrows represent the upstream edge modes. For fully pinched SG (at $V_{SG}=-2$ V), all the edge modes move between the floating contacts (Fig.~\ref{fig3}(c)). The outermost integer modes are directed to the ground, with the outer at $V_{SG}=-1$ V (Fig.~\ref{fig3}(d)) and the inner at $V_{SG}-0.52$~V (Fig.~\ref{fig3}(e)), leaving only the fractional modes (red and blue dashed arrows) propagating between the floating contacts. Finally, with the SG fully open (Fig.~\ref{fig3}(f)), all the edge modes end up in the ground.
\begin{center}
\begin{table*}[ht]
\renewcommand{\arraystretch}{1.0} 
\begin{tabular}{c|p{4cm}|c|c|c|c}
 $\nu$ & $N$ & $k_d(N)$ & $k_u(N)$& $k_{xy}(N)$ & $k_{bulk}$\\ 
 \hline
\multirow{3}{2em}{5/2} & all modes, $N=3$ & $k_d(3)=3.3k_0\pm0.05k_0$ & \multirow{3}{8em}{$\sim0.8k_0\pm0.02k_0$} &$k_{xy}(3)=2.5k_0\pm0.07k_0$ & \multirow{3}{8em}{$\sim0.54k_0\pm0.03k_0$}\\ 
& inner modes, $N=2$ & $k_d(2)=2.19k_0\pm0.03k_0$ & &$k_{xy}(2)=1.39k_0\pm0.05k_0$\\
& innermost fractional modes, $N=1$ & $k_d(1)=1.09k_0\pm0.02k_0$ & &$k_{xy}(1)=0.29k_0\pm0.04k_0$\\
\hline
\multirow{4}{2em}{7/2} & all modes, $N=4$ & $k_d(4)=4k_0\pm0.05k_0$ & \multirow{3}{8em}{$\sim0.52k_0\pm0.02k_0$} &$k_{xy}(4)=3.48k_0\pm0.07k_0$ & \multirow{3}{8em}{$\sim0.6k_0\pm0.04k_0$}\\ 
& inner modes, $N=3$ & $k_d(3)=2.92k_0\pm0.07k_0$ & &$k_{xy}(3)=2.4k_0\pm0.05k_0$\\
& inner two modes, $N=2$ & $k_d(2)=1.51k_0\pm0.03k_0$ & &$k_{xy}(2)=1k_0\pm0.05k_0$\\
& innermost fractional modes, $N=1$ & $k_d(1)=0.74k_0\pm0.04k_0$ & &$k_{xy}(1)=0.22k_0\pm0.04k_0$\\
\hline
\end{tabular}
\caption{\label{Table1}Summary of measured downstream and upstream thermal conductance coefficients $k_d(N)$ and $k_u(N)$ and topological thermal Hall conductance coefficient $k_{xy}(N)$ as a function of the number of downstream charged edge modes ($N$) between the floating contacts L and R for the $\nu=5/2$ and $\nu=7/2$ states. $k_{bulk}$ is the bulk thermal conductance.}
\end{table*}
\end{center}
Table.~\ref{Table1} outlines the thermal conductance derived from the plots in Fig.~\ref{fig3}(a) and \ref{fig3}(b). Here, the coefficients $k_d(N)$ and $k_u(N)$ represent downstream and upstream thermal conductance coefficients, respectively for $ N$ number of downstream charged modes (integer and fractional) (\textbf{SM-3}). Accordingly, the topological thermal Hall conductance coefficient is $k_{xy}(N)=k_d(N)-k_u(N)$. The bulk \textcolor{black}{thermal conductance measured with the gate fully open} is $k_{bulk}\sim0.54k_0\pm0.03k_0$, (Fig.~\ref{fig3}(b)). Interestingly, we find a half-integer value, $k_{xy}(3)=2.5k_0\pm0.07k_0$, with SG fully pinched (Fig.~\ref{fig3}(c)). However, upon removing the outer integer edges, $k_{xy}$ (for $N=1,2$) deviates from half-integer values (see Table.~\ref{Table1}). Unlike at $\nu=2$, here, $k_{xy}$ decreases by $\sim1.1k_0$ (instead of $1k_0$) when the integer edge modes are removed, suggesting a greater energy loss of the inner modes when isolated. We will return to this issue later. 

For the second even-denominator $\nu=7/2$ state, we have measured a different device with an L to R distance of 30~$\mu$m and showing improved quality of the state (refer to \textbf{SM~Fig.1e} and \textbf{SM~Fig.1f}). The $J_{down}$ (or $J_{up}$) vs. $T_L^2-T_0^2$ (or $T_R^2-T_0^2$) plots corresponding to this state are presented in \textbf{SM-6}.
Table.~\ref{Table1} also summarizes the thermal conductance coefficients determined from these plots. Similar to $\nu=5/2$, here, we once again observe a half-integer value, $k_{xy}(4)=3.48k_0\pm0.07k_0$, with SG fully pinched and deviations from half-integer values with the outer edge modes removed. Note that the $\nu=7/2$ state has an extra integer mode compared to the $\nu=5/2$ state. \textcolor{black}{~However, unlike at $\nu=5/2$ we observe identical bulk and upstream thermal conductances: $k_{bulk}=0.6k_0\pm0.04k_0$ and $k_u\sim0.52k_0\pm0.02k_0$ (used to calculate $k_{xy}$). Note that $k_{bulk}$ is measured with  SG fully opened, while $k_u$ is measured with the SG closed, either fully or partially. Thus, the observed identical values indicate complete heat loss from the upstream edges, consistent with the device's longer floating contact separation (30 $\mu m$).}

The observed half-integer $k_{xy}$ values for fully closed SG align with the PH-Pf order. In \textbf{SM-7}, we compare the measured $k_{xy}$ for different numbers of edge modes with the expected values for the PH-Pf order. The comparison shows a lower than expected $k_{xy}$ of the most inner channel when the integer modes are peeled away for both the studied states. It is crucial to note that while the measurement of $k_{2t}$ relies on the heat that leaves the heated source contact, our present measurement of $k_d$ or $k_u$ depends on the heat that reaches the contact R (or, alternatively the contact L), which may suffer from heat loss on its way (after emanating from the heated contact). We find that the measured upstream heat conductance $k_u$ is independent of gate voltage (refer to \textbf{SM-5}), suggesting constant heat loss (possibly to the bulk). Consequently, the observed deviation in $k_{xy}$ indicates increased heat loss from the inner downstream charged modes. We suggest that as we peel off the integer modes by `opening' the SG gate, the confining potential gets softer, leading to a low drift velocity of the edge modes and thus increasing its heat loss.\textcolor{black}{~Additionally, partial equilibration among all modes in the un-gated etched regions, each $\sim 2um$ long, between SG and each floating contact can alter the total heat transported by the edge modes propagating along the partially closed gate. This might also lead to unequal heat losses in the upstream and downstream directions, thus contributing to the deviation in $k_{xy}$ in these conditions. These handicaps were considered in future designs with shorter etched regions.}

\section{Conclusion}
This rather elaborate measurement configuration was developed to obtain the topological order of the states in the FQH regime~\cite{melcer2023direct,melcer2024heat}. This method overcomes the drawback of previously tested `two-terminal' measurements (the possible lack of full thermal equilibration)~\cite{banerjee2018observation,dutta2022isolated,jezouin2013quantum}. It allows employing higher-quality 2DEG (with wider conductance plateaus and lower longitudinal resistance) by eliminating the contribution to heat transfer of the highly intricate doping configuration~\cite{banerjee2018observation,melcer2023direct,umansky2009mbe,umansky2013mbe}. While the long-standing theoretical prediction of the order of the even-denominator states is the A-Pf order~\cite{morf1998transition,levin2007particle,lee2007particle}, all our measurements indicate the PH-Pf order~\cite{banerjee2018observation,dutta2022isolated,dutta2022distinguishing}. Finding the same order in the two even denominator states may point to a more fundamental reason for the consistent finding of the PH-Pf order. It will be commendable if new theoretical works address the unavoidable disorder and Landau-level mixing in the GaAs devices~\cite{storni2010fractional,wojs2010landau,zucker2016stabilization,das2023anomalous,herviou2023possible}. These shortfalls might be overcome in high-quality hBN-encapsulated graphene hosting even denominator states~\cite{huang2022valley}.

\section{Acknowledgement}
We thank Dima E. Feldman for the fruitful discussions. We also thank Sourav Manna and Ankur Das for the valuable discussions and suggestions. M.H. acknowledges the support of the European Research Council under the European Union's Horizon 2020 research and innovation program (grant agreement number 833078).\\

$^{\dagger}$ Contributed equally\\
$^{\star}$ Corresponding author: \textcolor{red}{moty.heiblum@weizmann.ac.il}
%
\clearpage


\section*{Supplementary material}

# Supplemental Material

# Topological Thermal Hall Conductance of Even Denominator Fractional States

Arup Kumar Paul[1†], Priya Tiwari[1†], Ron Melcer[2], Vladimir Umansky[1], and Moty Heiblum[1*]

[1] Braun Center for Submicron Research, Department of Condensed Matter Physics,
Weizmann Institute of Science, Rehovot, Israel

[2] Qedma Quantum Computing, Tel Aviv, Israel

**Supplemental Material-1: Quantum Hall response of the thermal conductance devices**

Supplemental Material (SM) Figure 1a shows $R_{xy}$ (in red) and $R_{xx}$ (in blue) as functions of magnetic field $B$ around the ν=5/2 and ν=7/2 (inset) plateaus in the Hall bar sample. Unlike the Hall bar, the thermal conductance device's geometry doesn't permit direct measurement of the longitudinal resistance $R_{xx}$ [1][2]. Instead, we rely on bulk leakage current ($\Delta I$) between the floating contacts (L and R) to identify the minima of bulk conductance $G_{bulk}$ (equivalent to minima of $R_{xx}$). SM Figure 1b illustrates the low-frequency (~17Hz) setup to measure $G_{bulk}$. An AC current $I_{ac}$ is injected from source contact S1 and in the absence of any bulk leakage, it is carried to the floating contact L by the edge modes shown by the red arrows. From the floating contact, the current splits between the outgoing edge modes (colored orange). The voltage ($V_L$) of the floating contact L is related to the outgoing currents $I_{out}$ in each mesa-arm by,

$$V_L = I_{out}R_H \tag{S1}$$

were $R_H$ being the quantum Hall (QH) resistance. The current $I_{out}$ in every arm (thus $V_L$) can be determined from the voltage drop ($V_m$) on separate ohmic contacts that connect the outgoing edge modes to the ground, like the voltage probe contact V1 in the top-left arm, using:

$$V_m = I_{out}R_H = V_L \tag{S2}$$

We found $I_{out} \sim \frac{I_{ac}}{3}$ for all the arms at the QH plateau centers, even for the ν=7/2 and ν=5/2 states, which have finite bulk conductance. The equal splitting of the injected current indicates significantly small or absent bulk conductance at the plateau centers. Also,

$$V_L \sim \frac{I_{ac}}{3}R_H \tag{S3}$$

---

[†]These authors contributed equally
[*]Corresponding author: moty.heiblum@weizmann.ac.il

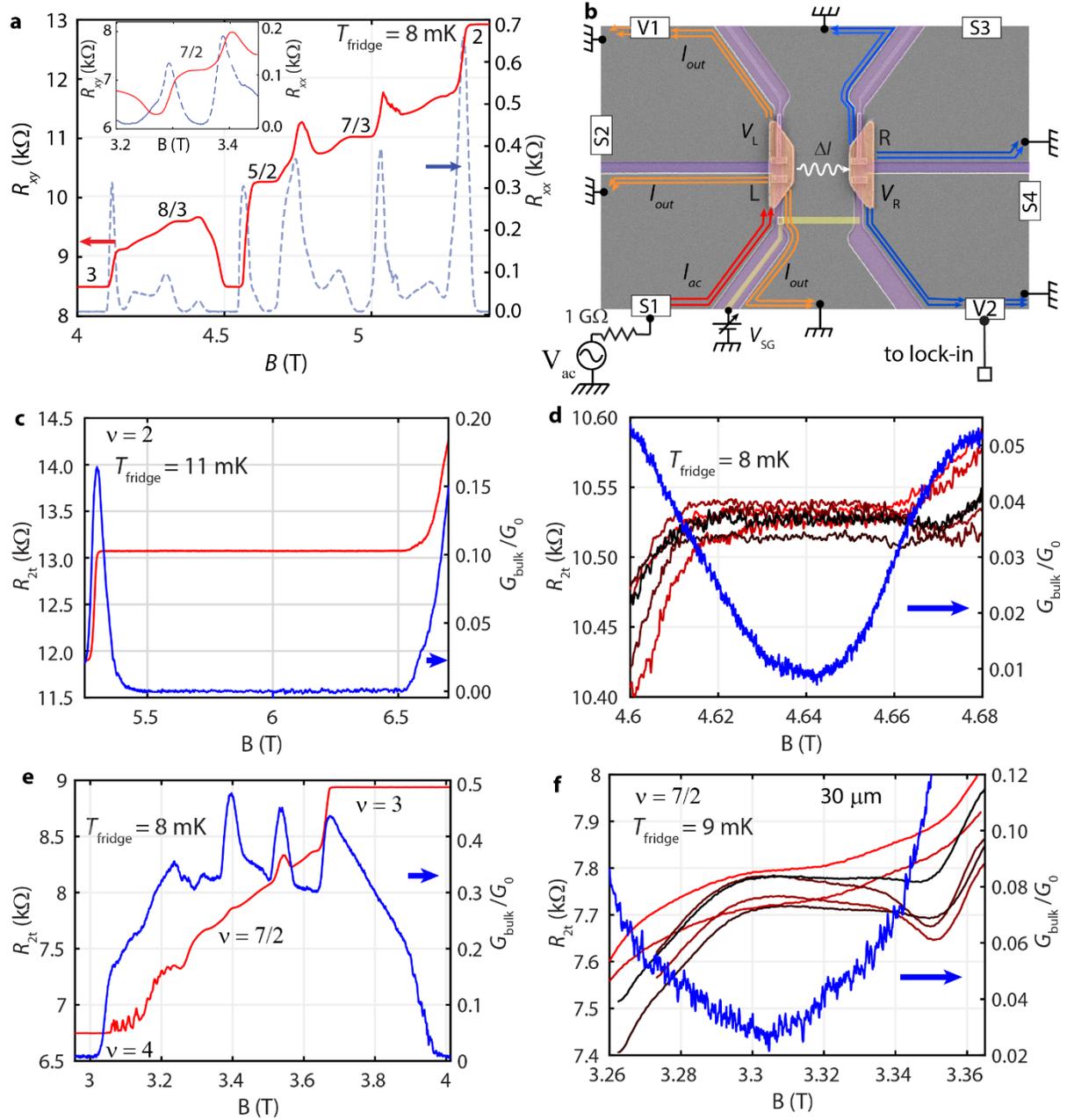

**Supplemental Material Figure 1: Quantum Hall responses - (a)** Quantum Hall response around the ν=5/2 and the ν=7/2 (inset) states in the Hall bar device. **(b)** Measurement configuration for determining bulk conductance $G_{bulk}$ between the floating contacts. The gate SG is fully open, diverting all outgoing edge modes from contact L to ground. **(c)-(e)** Two-terminal quantum Hall resistance $R_{2t}$ and $G_{bulk}$ as functions of the applied field B at ν=2, ν=5/2, and ν=7/2 in the device with 10μm separation between the floating contacts. $R_{2t}$ traces measured at different contacts are shown by the colors red to black. The measured $G_{bulk}$ is depicted in blue. $G_0$ is the conductance quanta. **(f)** $R_{2t}$ and $G_{bulk}$ vs. B responses around the ν=7/2 state in the device with a 30μm separation between the floating contacts.

Similarly, in the presence of bulk leakage, any small current $\Delta I$ reaching the other floating contact R creates a voltage drop,

$$V_R \sim \frac{\Delta I}{3} R_H \qquad (S4)$$

This voltage is measured at the contact V2 in the bottom-right mesa arm. The leakage current $\Delta I$ depends on the difference, $V_L$-$V_R$ and bulk resistance ($R_{bulk}$), between the floating contacts:

$$V_L\text{-}V_R \sim \Delta I \, R_{bulk} \qquad (S5)$$

Using **Eqs. S3** to **S5** we get,

$$R_{bulk} = (R_H/3)(V_L\text{-}V_R)/V_R \sim (R_H/3)(V_L/V_R) \sim$$

$$\sim \left(\frac{R_H^2}{9}\right)\left(\frac{I_{ac}}{V_R}\right), \qquad (S6)$$

Here, we have ignored the second term on the right-hand side since the measured $V_L$ is two orders larger than $V_R$. Finally, **Eq. S6** leads to,

$$G_{bulk} = \frac{1}{R_{bulk}} = \frac{9}{R_H^2}\left(\frac{V_R}{I_{ac}}\right) \qquad (S7)$$

The SM Figure 1c, 1d, and 1e display the measured $G_{bulk}$ (in blue) as a function of the magnetic field ($B$) for filling factors $\nu$=2, $\nu$=5/2, and $\nu$=7/2, respectively in the device with 10μm L-R separation. Additionally, two-terminal Hall resistance ($R_{2t}$) traces (in red to black) measured at different voltage probes or current injection contacts in the sample are also shown. $R_{2t}$ is determined by measuring the voltage drop at the contacts while injecting a known current. For the integer $\nu$=2 state, $G_{bulk}$ remains zero throughout the plateau (SM Figure 1c). In the $\nu$=5/2 state, $R_{2t}$ traces exhibit clear plateaus for different contacts (SM Figure 1d). The $G_{bulk}$ reaches a minimum of $\sim$0.01$G_0$, near the plateau's center at $B$=4.64T, while increasing rapidly away from the center of the plateau. The $\nu$=7/2 state, however, is not well developed, as seen from the $R_{2t}$ and $G_{bulk}$ responses in SM Figure 1e. The SM Figure 1f shows the $\nu$=7/2 state's response in the device with 30μm separation between the floating contacts. In this device, the $\nu$=7/2 state is well developed, with $R_{2t}$ plateaus visible in different contacts. Here, $G_{bulk}$ shows a clear minimum of $\sim$0.03$G_0$ at 3.305 T, prompting us to check the thermal conductance of $\nu$=7/2 state in this device.

**Supplemental Material-2: Thermal conductance measurement**

There are two steps in the thermal conductance measurement [1][2]. **Step 1** is the measurement of the power-meter (PM) temperature using Johnson-Nyquist (J-N) thermal noise when the source floating contact (S) is being heated. In **Step 2**, we use the PM temperatures measured in Step 1 to determine the arriving power using a temperature vs.

input power response of the PM. The following sections describe these two steps using the downstream thermal conductance measurement at ν=2 as an example. Note that, for the upstream case, the roles of the floating contacts are reversed.

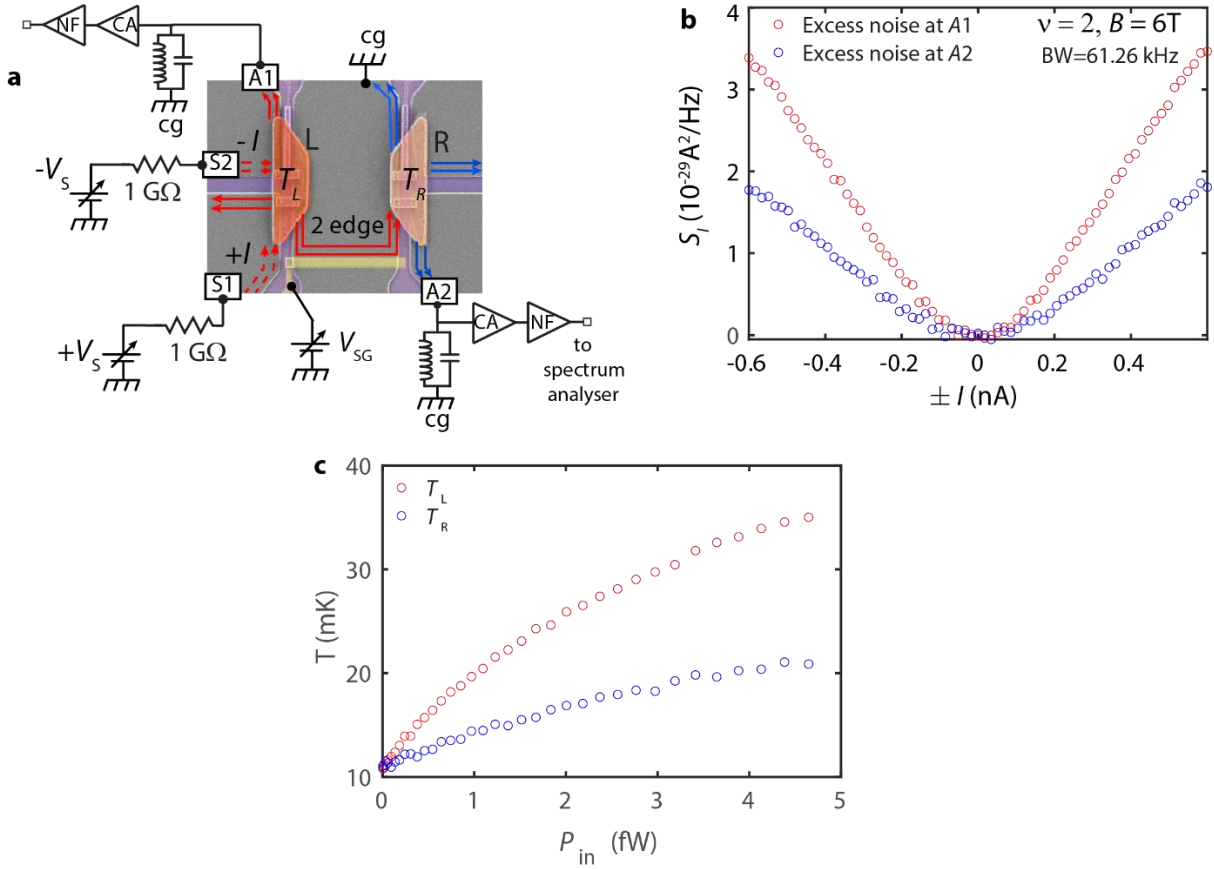

**Supplemental Material Figure 2: Excess thermal noise measurement - (a)** Setup to measure excess thermal noise from the floating contacts: the contact L (dark red) heated with known electrical power acts as the temperature source, while the contact R (light red) heats up by arriving power from the source, thus acts as a power meter (PM). **(b)** Measured excess noise ($S_I$) from the floating contacts plotted against the injected current ($I$) to the source. The colors red and blue are for the source (contact L) and power meter (contact R) noises, respectively. **(c)** Temperatures $T_L$ (red) and $T_R$ (blue) of the floating contacts determined from $S_I$ using **Eq. S8** and **Eq. S9**, as a function of the applied power $P_{in}$ to the source (contact L).

**Step 1**: In this step, one of the floating contacts is directly heated with a known electrical power, functioning as a *temperature source* (S). The other floating contact labeled as the *power meter* (PM) heats up due to incoming power carried by the arriving edge modes or the bulk from the temperature source. The heating of both contacts results in outgoing edge

modes carrying excess J-N noise, subsequently converted to the temperature of the floating contacts.

SM Figure 2a shows the setup where the floating contacts L and R act as the *source* and PM, respectively. Contact L is heated with the equal and opposite currents (dc) *+I* and *-I* injected from contacts S1 and S2. These currents are carried by the edge modes denoted as red dashed arrows. The DC currents are injected using the 1GΩ resistances in series with the sample and the DC voltage sources $V_S$. The injected currents cause net power dissipation $I^2R_H$ at the floating contact L. At steady state, the dissipated power equals the net outgoing power carried by the outgoing edge modes (red solid lines). By applying a suitable gate voltage $V_{SG}$, the edge modes in the middle arm are directed to contact R, increasing its temperature. The resulting excess J-N thermal noises ($S_I$) are measured by amplifiers connected to the contacts A1 and A2 for L and R, respectively. The amplifier lines play as an LC tank circuit, followed by a cold amplifier (CA) at a temperature 4K, followed by a room temperature (NF) amplifier. Noise is measured at the LC resonance frequencies (~632 kHz and ~717 kHz for L and R, respectively), with a bandwidth (BW) of ~61.3kHz. SM Figure 2b presents the measured excess J-N noise $S_I$ from the two floating contacts as a function of the injected current *I* when contact L is heated. SM Figure 2c presents the excess noise being converted to the temperature of the floating contacts ($T_L$ and $T_R$) as a function of dissipated power ($P_{in}=I^2R_H$). We use the following expressions for the conversion [1][2]:

$$S_I = 2k_B \Delta T \, G^*  \qquad (S8)$$

where $k_B$ is the Boltzmann constant, $\Delta T$ ($T_L$-$T_0$ or $T_R$-$T_0$) is the excess temperature of the floating contacts and $G^*$ is given by:

$$G^* = \frac{G_{S \to A} G_{S \to G}}{G_{S \to A} + G_{S \to G}} \qquad (S9)$$

with $G_{S \to A}$ is conductance between source and amplifier contact, and $G_{S \to G}$ is the conductance between the source and cold grounds.

**Step 2:** Note that in the previous step, we have determined PM temperatures from the excess noise. However, the net arriving power which caused the PM's temperature to rise is not yet known. In this step, we use the PM temperatures to determine the corresponding arriving powers. For this the PM is heated with known electrical power to establish its temperature vs. input power response, labeled as the calibration curve. Leveraging this curve, the PM temperature ($T_R$) from the preceding step is translated to arriving power. With the exception of heating the PM instead of the source, the rest of the calibration step is identical to step 1.

SM Figure 3a illustrates the setup for generating the calibration curve for contact R. In this configuration, the contact R (rather than L) is heated with equal and opposite DC currents

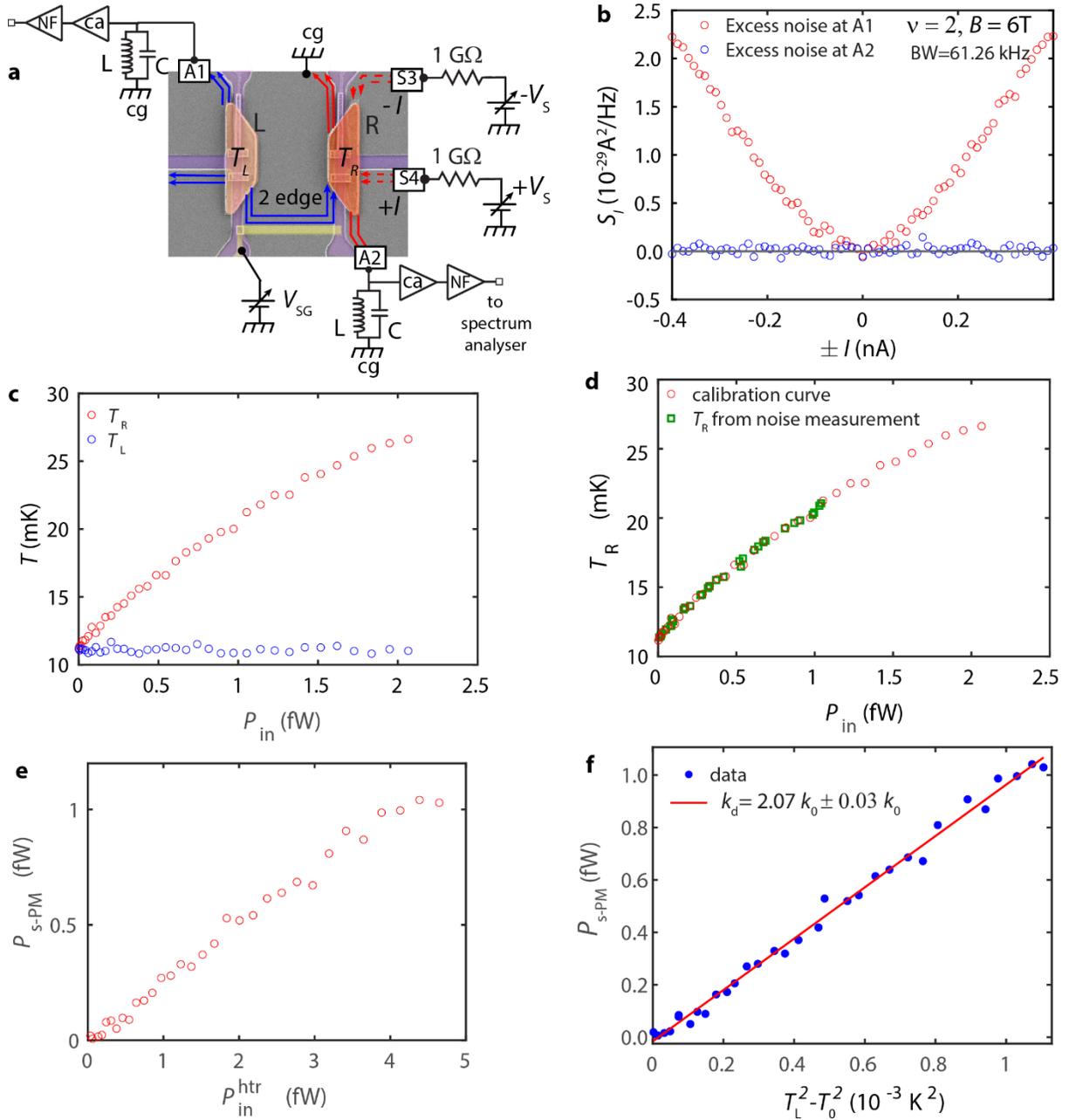

**Supplemental Material Figure 3: Calibration of power-meter-(a)** Setup to establish the temperature vs. input power ($P_{in}$) calibration curve for the PM (contact R, dark red). **(b)** Excess noise ($S_I$) from the floating contacts plotted against the injected current ($I$) to the PM contact R. The colors red and blue are for excess noises from contacts R and L, respectively. **(c)** Temperatures $T_R$ (red) and $T_L$ (blue) of the floating contacts were determined from $S_I$ as a function of the injected power $P_{in}$ to the PM. The $T_R$ vs. $P_{in}$ curve serves as the calibration curve. **(d)** The calibration curve (red) together with the PM temperatures (green square) from step 1 to determine the arriving power ($P_{s-PM}$) to PM when the source (contact L) is heated. **(e)** $P_{s-PM}$ against the power injected $P_{in}^s$ to source in step 1. **(f)** $P_{s-PM}$ (blue) as function of $T_L^2 - T_0^2$ with $T_L$ the temperature of the source contact, for determining the thermal conductance. Red solid line is the fit to determine downstream thermal conductance $k_d$.

6

*+I* and *-I*, injected via the source contacts S3 and S4, respectively. The excess noise ($S_I$) for both floating contacts, resulting from the heating of contact R, are presented in <span style="color:red">SM Figure 3b</span>. Like the previous step, the excess noise responses are converted to temperatures of the floating contacts using **Eqs. S8** and **S9**. <span style="color:red">SM Figure 3c</span> shows the resultant temperatures $T_R$ and $T_L$ as functions of the input power $P_{in}$ to contact R. Here, the $T_R$ vs. $P_{in}$ response represents the calibration curve. We use this calibration curve to determine the input power to PM corresponding to the measured PM temperatures when the source (contact L) is heated. In <span style="color:red">SM Figure 3d</span>, the calibration curve is depicted in red, with the green squares representing the measured PM (contact R) temperatures in Step 1. The power arriving at the PM ($P_{\text{S-PM}}$, from source) corresponding to the green squares is then plotted against input power $P_{in}^s$ to the source L (in step 1) in <span style="color:red">SM Figure 3e</span>. Finally, <span style="color:red">SM Figure 3f</span> shows the $P_{\text{S-PM}}$ vs. $T_L^2 - T_0^2$ plot to determine the thermal conductance. Here, $T_L$ is the source temperature in step 1.

### Supplemental Material-3: Controlling number of edge modes between the floating contacts

We use the gate SG to control the number of edge modes between the floating contacts. To select appropriate gate voltages ($V_{SG}$), we utilize the gate transmittance ($t$) versus $V_{SG}$ responses. <span style="color:red">SM Figure 4a</span> illustrates the measurement configuration to find the transmittance $t$ as a function of $V_{SG}$. The gate SG (yellow-shaded region) divides the middle mesa into three parts: the upper part between floating contacts (L and R), the mesa directly below SG, and the lower part. By adjusting the filling factor below SG with $V_{SG}$, downstream edge modes (blue arrows) from the left floating contact L can be directed to the bottom cold-ground (cg) or to the right floating contact R. To measure $t$, a low-frequency ac current $I_{in}$ is injected from the contact S5 (in the lower part). The current is carried by downstream edge modes (red arrows) to the SG-controlled region. From here, the reflected edges reach the cold ground. The transmitted edges enter floating contact R, which splits the transmitted current $I_t$ equally between outgoing edge modes (orange arrows). $I_t$ is measured from the voltage drop $V_m$ at the contact V2 using:

$$V_m = \frac{I_t}{3} R_H \qquad (S10)$$

where $R_H$ is the QH resistance. Consequently, $t$ is given by:

$$t = \frac{I_t}{I_{in}} = \frac{3V_m}{I_{in} R_H} \qquad (S11)$$

In this configuration, the number ($N$) of downstream edge modes moving from L to R equals the number of reflected downstream edges at SG. Therefore, $t$ can be translated into effective filling factor ($\nu_{LR}$) of the edge modes moving from L to R, with

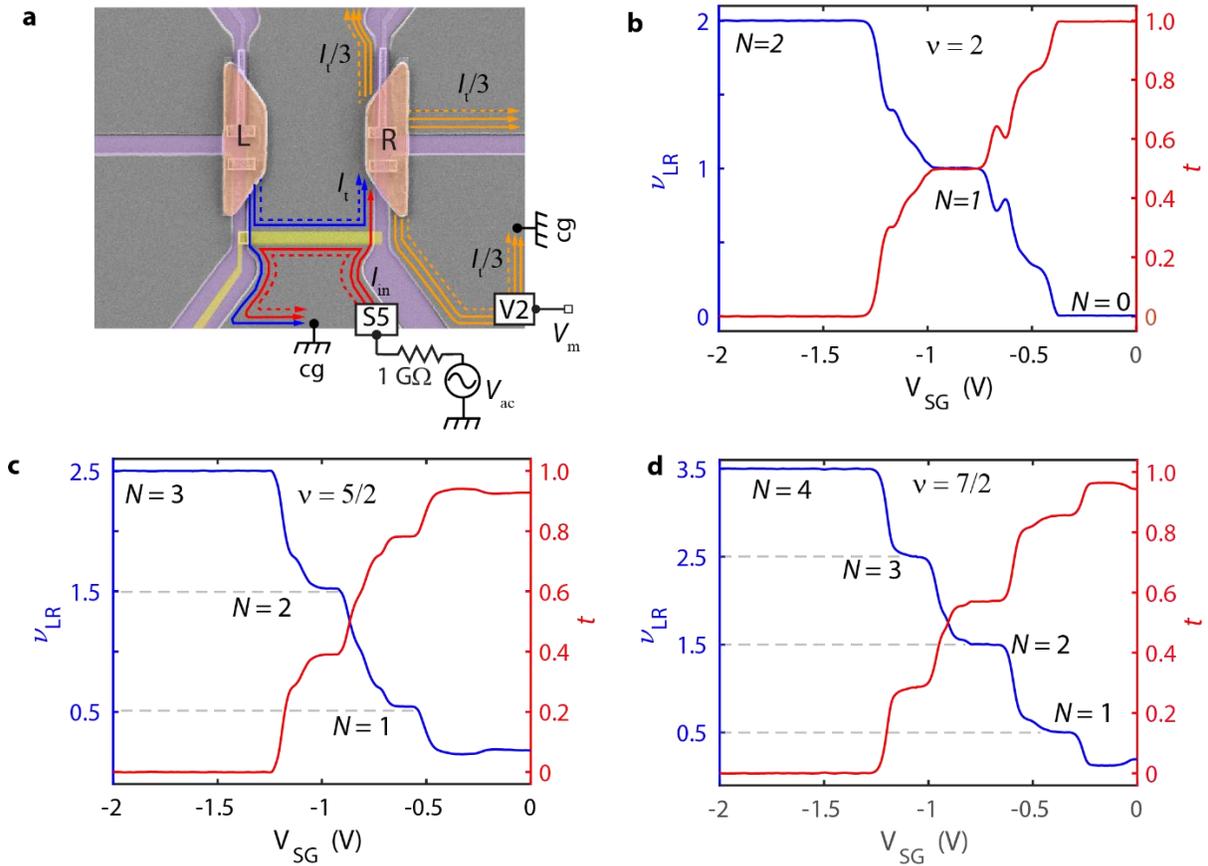

**Supplemental Material Figure 4: Gate responses - (a)** Measurement setup for gate transmittance (*t*) versus gate voltage (*V*$_{SG}$) responses. **(b), (c)** and **(d)** Transmittance (*t*, in red) and corresponding ν$_{LR}$ (in blue) as a function of the gate voltage *V*$_{SG}$ for the ν = 2, 5/2 and 7/2 states, respectively. The horizontal dashed lines mark the plateaus at ½ integer values. The edge mode number *N* corresponding to the plateaus are shown in the figures.

$$\nu_{LR} = (1\text{-}t)\nu \qquad\qquad (S12)$$

where ν is the global filling factor set by applied magnetic field *B*. Note, ν$_{LR}$ determine the effective electrical conductance $G_{LR}$= ν$_{LR}G_0$ between the floating contacts. For *t*=0, the gate is fully pinched (ν$_{LR}$ = ν), allowing all available modes to propagate between floating contacts. Conversely, at *t*=1, the gate is fully open ( ν$_{LR}$ = 0), preventing any direct edge propagation between floating contacts.

SM Figure 4b to 4d depicts measured *t* (in red) and ν$_{LR}$ (in blue) as functions of the gate voltage *V*$_{SG}$ for the studied filling factors. For ν = 2 we find ν$_{LR}$ = 0 (i.e. *t*=1), for *V*$_{SG}$ > - 0.38 V (SM Figure 4b), indicating a fully open gate. For the even-denominator states, the gate

remains fully open at $V_{SG}$=0 before applying any non-zero gate voltage. However, once a voltage is applied, the gate does not fully open again at $V_{SG}$=0. Consequently, the transmittance never returns to unity, (see <span style="color:red">SM Figure 4c</span> and <span style="color:red">SM Figure 4d</span>). This behavior is due to the hysteretic nature of the complicated SPSL-type doping in our GaAs samples. Regardless, in all cases, as $V_{SG}$ decreases, $t$ approaches zero after consistently passing through multiple (single for $\nu = 2$) plateaus. As a result, $\nu_{LR}$ increases and eventually match the actual filling factor $\nu$ (beyond $V_{SG}$ < -1.3V). These plateaus signify the sequential addition of the edge modes between L and R. Therefore, to vary the number of edge mode ($N$) for thermal conductance measurement, gate voltages are set in the middle of these plateaus. As seen, for $\nu = 2$, a single intermediate plateau appears at $\nu_{LR} = 1$ ($t = 0.5$), indicating the presence of a single integer edge mode ($N$=1) between L and R. For the even denominator states, the plateaus appear at half-integer values of $\nu_{LR}$, as marked by the horizontal dashed lines in <span style="color:red">SM Figure 4c</span> and <span style="color:red">SM Figure 4d</span>. The plateau with $\nu_{LR} = 0.5$ is for the isolated fractional ½ charged mode ($N$=1) with $G_{LR}=\frac{1}{2}G_0$. The subsequent plateaus appear at $\nu_{LR} = 1.5$, 2.5 and 3.5 ( $\nu_{LR} = 3.5$ is exclusive to $\nu = 7/2$ state), indicating the inclusion of one, two and three integer edge modes, respectively, atop the ½ fractional mode (resulting in $N$=2,3 and 4, respectively).

**Supplemental Material-4: $k_{xy}$ in presence of bulk heat conductance at $\nu$=2**

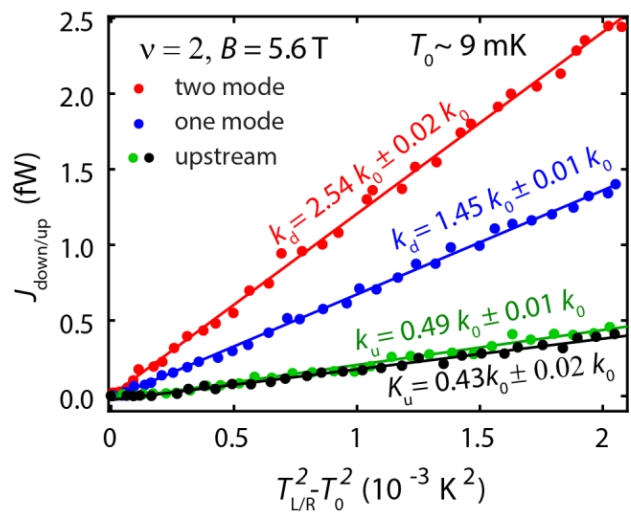

**Supplemental Material Figure 5: Thermal conductance in presence bulk-** Downstream (or upstream) heat current $J_{down}$ (or $J_{up}$) vs. $T_L^2 - T_0^2$ (or $T_R^2 - T_0^2$) at $\nu$=2, but away from the center of the plateau at B=5.6 T. The solid lines are linear fit to the plot to find the thermal conductance (color coded) $k_d$ (or $k_u$). The black and green colors represent net upstream heat flow when only the inner downstream mode (one mode) and both the downstream edge modes (two modes) move between the floating contacts.

As mentioned in the main text, to assess the effect of bulk heat conductance on $K_{xy}$, we have measured $J_{down}$ and $J_{up}$, by tuning the magnetic field away from the center of the ν=2 plateau. In the presence of bulk heat conductance and when the gate SG is closed (fully or partially), the net downstream (or the upstream) heat flow includes heat flow via edge modes and the bulk. Thus, increased heat conductance is expected. SM Figure 5 shows $J_{down}$ ($J_{up}$) versus $T_L^2$-$T_0^2$ ($T_R^2$-$T_0^2$) plots measured at 5.6T. At this field the QH bulk is electrically insulating (SM Figure 1c), however, we see bulk heat conduction. Here, we find $k_d$ = 2.54$k_0$ ± 0.02$k_0$ for two edge modes (red circles) and $k_d$ = 1.45 $k_0$ ± 0.01 $k_0$ for a single mode (blue circles), both increased compared to the measured values at the center of the plateau. Also, for upstream heat flow, we measured non-zero thermal conductance $k_u$ = 0.43 $k_0$ ± 0.02 $k_0$ (two modes, green circles) and $k_u$ = 0.49 $k_0$ ± 0.01 $k_0$ (one mode, black circles), unlike at the center of the plateau. The increase shows additional heat conduction via bulk. Nevertheless, we find $k_{xy}$ = 2.11 $k_0$ ± 0.03 $k_0$ and $k_{xy}$ = 0.96 $k_0$ ± 0.02 $k_0$ for two and one modes, respectively. Thus, the method provides accurate topological thermal Hall conductance, even with bulk heat flow. Note that for both upstream and downstream, the increase in thermal conductance is around ∼ 0.5 $k_0$, suggesting no heat loss from the edge modes.

**Supplemental Material-5: Upstream thermal conductance**

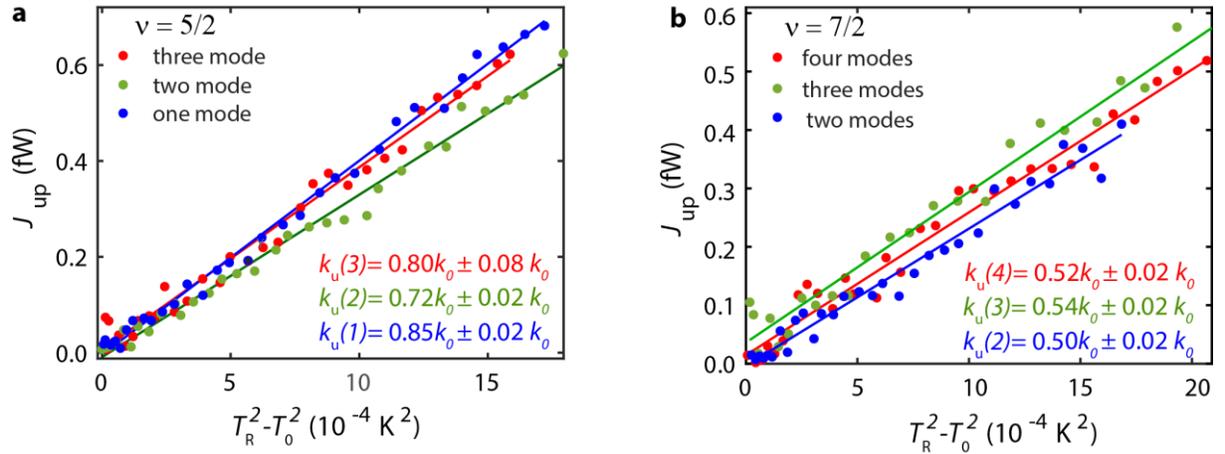

**Supplemental Material Figure 6: Upstream thermal conductance - (a)** and **(b)** Upstream heat current $J_{up}$ vs. $T_R^2 - T_0^2$ plots for different number of edge modes ($N$) between the floating contacts at ν =5/2 and ν =7/2 fillings, respectively. The solid lines are linear fit to the plot to find the thermal conductance $k_u(N)$. The $k_u(N)$ values determined from the plots are color coded.

For upstream heat conductance ($k_u$) measurement, we reverse the roles of the floating contacts: the right floating contact R becomes the *temperature source*, and the left contact L serves as the *power-meter* (PM). SM Figures 6a and 6b show the upstream heat current $J_{up}$ relative to $T_R^2 - T_0^2$ for different edge mode numbers ($N$) between L and R, for the $\nu=5/2$ and $\nu=7/2$ states, respectively. The $k_u(N)$ values (color-coded) determined from the plots are also shown in the figures. As can be seen, for both the states, $k_u$ is independent of $N$. Also, for $\nu=5/2$, the measured upstream conductance $k_u \sim 0.8\ k_0$ is more than the bulk thermal conductance $k_{bulk} \sim 0.5\ k_0$ (see Figure 3b of the main text), suggesting additional $\sim 0.3\ k_0$ is coming from the upstream Majorana mode. It can be seen that the contribution of the Majorana mode is less than the expected $0.5\ k_0$, which is most likely due to heat loss to the bulk or downstream charged modes. We discuss the bulk and upstream heat flow for the $\nu=7/2$ state, in the next section.

**Supplemental Material-6: Thermal conductance at $\nu=7/2$**

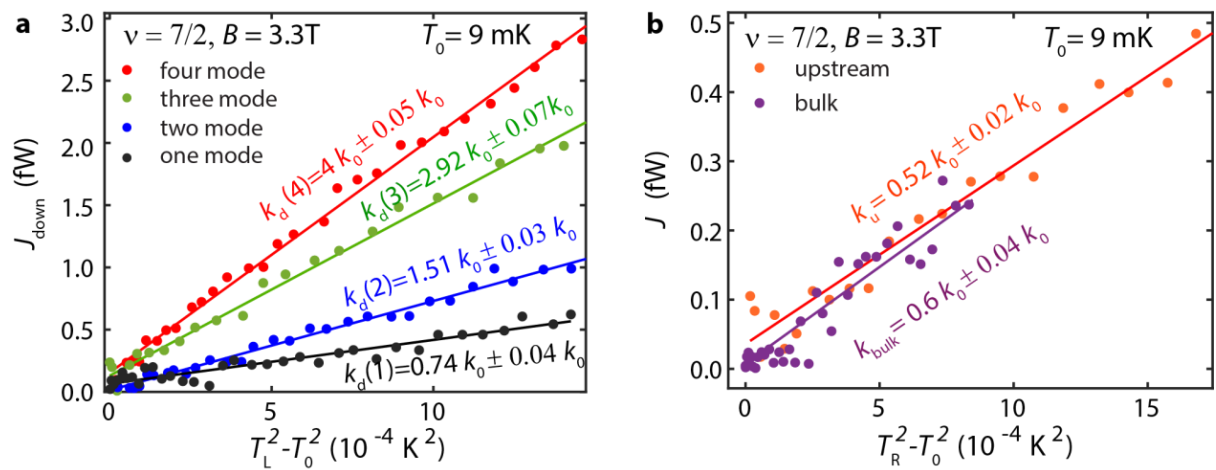

**Supplemental Material Figure 7: (a)** $J_{down}$ vs. $T_L^2 - T_0^2$ plots at the center of $\nu = 7/2$ plateau at $\sim 3.3$ T ($T_0 \sim 9$ mK, electron temperature). The red, green, blue, and black data points correspond to all edge modes (three integers and the fractional mode), inner three (two integers and the fractional mode), inner two (innermost integer and the fractional mode), and innermost fractional mode, respectively. The corresponding downstream heat conductance coefficients ($k_d(N)$) are color-coded. **(b)** Upstream heat flow $J_{up}$ for fully pinched gate SG, ($N=4$) as a function of $T_R^2 - T_0^2$ (orange). The same plot for bulk heat transport, when SG is fully open, is shown in purple.

SM Figure 7a shows $J_{down}$ as a function of $T_L^2 - T_0^2$, measured at the center of the $\nu = 7/2$ plateau. Similar plots of $J_{up}$ and the bulk's contribution are shown in SM Figure 7b. We observe a slightly higher $k_{bulk} \sim 0.6\ k_0$, compared to the average upstream $k_u \sim 0.52\ k_0$, which is unusual. However, the plot for bulk heat conductance closely follows the plot for upstream heat

conductance. The observed higher value can be attributed to fitting error arising from more noisy data of bulk taken within a lower temperature range. By comparing the figures SM Figure 7b and SM Figure 6b, it can be seen that unlike at ν = 5/2, for the ν = 7/2 state, the measured bulk thermal conductance $k_{bulk}$ is similar to the upstream thermal conductance, indicating that all heat carried by the upstream Majorana mode is lost. This is consistent with the longer edge mode travel length (~30μm) in the device used for measuring the 7/2 state.

**Supplemental Material-7: $k_{xy}$ as a function of $N$**

SM Figure 8a and SM Figure 8b show measured $k_{xy}$ (red error bars) as a function of the number of downstream charged edge modes ($N$) propagating from L to R, at ν=5/2and ν=7/2, respectively. The blue squares are the expected values of $k_{xy}$ of the PH-Pf order. The error bars in the figures correspond to statistical error. Note the more significant deviations from the expected values at $N$ =1 for ν=5/2 and similar deviations at $N$ =1 and 2 for ν=7/2. In SM Figure. 8a, the arrows indicate edge modes: red solid and dashed lines are for integer and fraction ½ mode. The blue dashed shows the upstream Majorana modes.

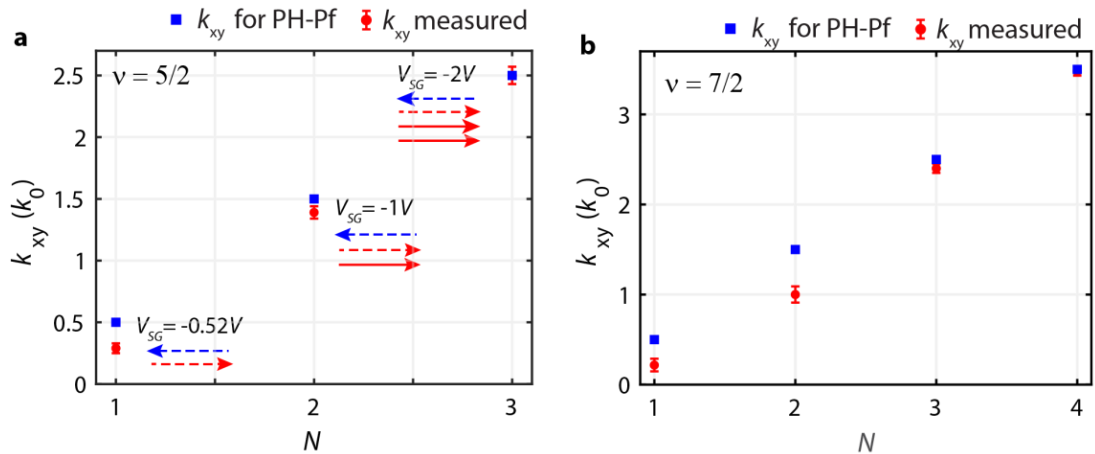

**Supplemental Material Figure 8: (a)** and **(b)** Measured $k_{xy}$ (red error bars) vs. number of downstream charged edge modes ($N$) at ν=5/2and ν=7/2, respectively. The expected values for the PH-Pf order are shown by blue squares.

**Supplemental Material-8: Magnetic field dependence of $k_{xy}$ at ν=5/2**

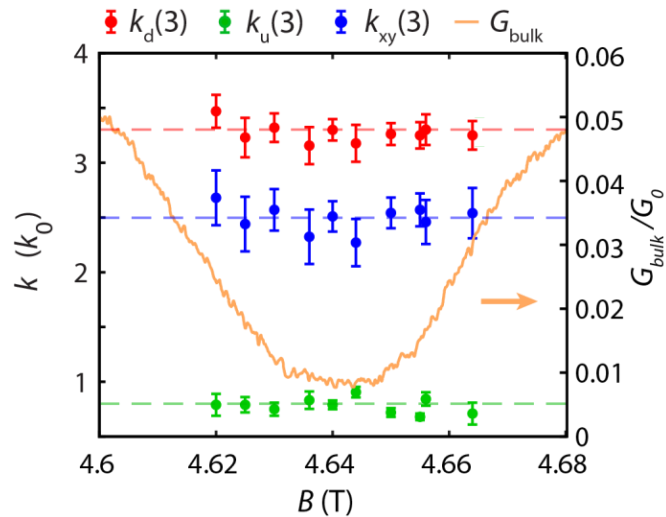

**Supplemental Material Figure 9:** bulk conductance $G_{bulk}$ (orange) between floating contacts, $k_{xy}$ (3) (blue), $k_d$ (3) (red), and $k_u$ (3) (green) - as a function of magnetic field (B) around the minima of $G_{bulk}$. The dashed lines show the mean values.

We repeated the $k_{xy}$ measurements at different locations on the ν=5/2 conductance plateau. The SM Figure 9 shows bulk conductance between floating contacts L and R, $G_{bulk}$ (orange) in units of $G_0$, $k_{xy}$ (3) (blue), $k_d$ (3) (red), and $k_u$ (3) (green) - as function magnetic field (B) around the minima of $G_{bulk}$. We find an average topological thermal conductance across the plateau, $K_{xy}$ (3) = $2.5k_0 \pm 0.2k_0$.

**Supplemental Material-9: Gain calibration of the cryo-amplifiers**

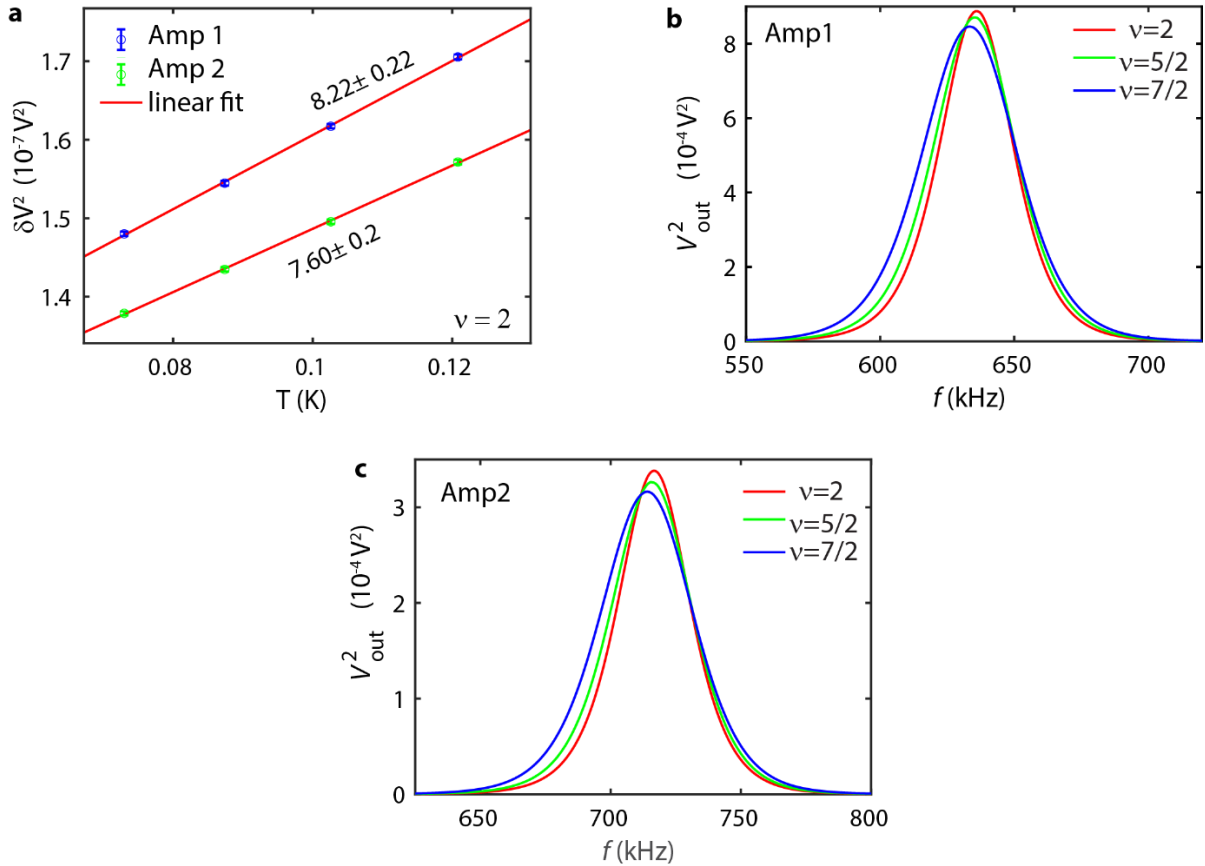

**Supplemental Material Figure 10: Gain calibration- (a)** Measured background voltage fluctuation $\delta V_{out}^2$ as a function of fridge temperature ($T$), at filling factor $\nu = 2$, for both amplifiers. Red lines are the linear fits to find the slopes to determine the amplifier gains. **(b)** and **(d)** Resonance plots at different filling factors for Amp1 and Amp2, respectively. The resonance frequencies are ~634 kHz and ~717 kHz, respectively. $V_{out}$ corresponds to the amplified output signal in response to a high-frequency signal applied to the sample.

The gains of the HEMT cryo-amplifiers (CA) vary with temperature cycle, necessitating gain calibration before conducting noise measurements. For the robust integer $\nu=2$ state we have used equilibrium Johnson-Nyquist noise to determine the amplifier gains. The equilibrium thermal noise at any temperature $T$ is given by $S_V = 4\,k_B TR = \langle\delta V_{th}^2\rangle/BW$, where $\delta V_{th}^2$ is thermal noise voltage fluctuation, $k_B$ is Boltzmann constant, $R$ is quantum Hall resistance, and $BW$ is the measurement bandwidth (~61.3 kHz). In SM Figure 10a, the total background noises $\left(\delta V_{out}^2\right)$ measured by the amplifiers are plotted against the fridge temperature $T$. The blue and green colors represent noises measured at the left (denoted as Amp1) and right (denoted as Amp2) amplifiers, respectively. The Amp1 and Amp2 are connected to the contacts A1 and A2, respectively, as shown in SM Figure 2 and SM Figure 3. The measured noise contains

14

amplified (gain $A$) thermal noise as well as the amplifier's own (current and voltage) noises ($\delta V_A^2$),

$$\delta V_{\text{out}}^2 = A^2 \ (\delta V_{\text{th}}^2 + \delta V_A^2) \tag{S13}$$

However, since the amplifiers sits at 4k, the amplifier noises are independent of the fridge temperature. This enables determination of the gain from the slope of the linear $\delta V_{out}^2$ vs, $T$ plots, using $A = \sqrt{slope/(4 \ k_B T \ R \ BW)}$. While measuring the 10 μm separation device, we measured gains 8.22±0.22, and 7.60±0.20 for Amp1 and Amp2, respectively, at ν=2. In the second measurement of the 30 μm separation device, Amp1 exhibited gain of 7.59±0.5, while Amp2 had gain of 5.03±0.11 at ν=2.

For the fragile even denominator states the amplifier gains are estimated by comparing the LC resonance plot areas associated to the states to that of at $\nu = 2$. This method utilizes the relation between the resonance plot area, amplifier gain and measurement bandwidth given by,

$$A = \frac{1}{V_{\text{in}}} \sqrt{\frac{Area}{BW}} \tag{S14}$$

where $V_{\text{in}}$ a frequency independent input signal to the amplifier contacts. Comparing resonance area with known gain to a resonance area for unknown gain eliminates $V_{\text{in}}$ which is difficult to measure accurately, due to capacitive losses. SM Figure 10b and 10c illustrates comparison of the resonance plots for both amplifiers at the studied filling factors. For the measurement of the $\nu = 5/2$ state, Amp1 gain was ~8.7, and Amp2 had gain of ~7.41. During the measurement of the $\nu = 7/2$ state, the gains were ~8.52 and ~5.64, respectively.

**Supplementary references:**


\clearpage
\end{document}